\documentclass[aps,prd,superscriptaddress,eqsecnum,nofootinbib,showkeys,twocolumn,showpacs,preprintnumbers]{revtex4-1}
\usepackage{graphicx}
\usepackage{euscript,amssymb}
\usepackage{amsfonts}
\usepackage{amsmath}
\usepackage{amssymb}
\usepackage{graphicx}
\usepackage{euscript}
\usepackage{color}
\begin{document}

\title{The cosmology of an holographic induced gravity model with curvature effects}

\author{Mariam Bouhmadi-L\'{o}pez}
\email{mariam.bouhmadi@ist.utl.pt}
\affiliation{Centro Multidisciplinar de Astrof\'{\i}sica - CENTRA, Departamento de F\'{\i}sica, Instituto Superior T\'ecnico, Av. Rovisco Pais 1,1049-001 Lisboa, Portugal}
\author{Ahmed Errahmani}\email{ahmederrahmani1@yahoo.fr}
\affiliation{Laboratory of Physics of Matter and Radiation, Mohammed I University, BP 717, Oujda, Morocco}
\author{Taoufiq Ouali}
\email{ouali\_ta@yahoo.fr}
\affiliation{Laboratory of Physics of Matter and Radiation, Mohammed I University, BP 717, Oujda, Morocco}

\begin{abstract}
We present an holographic model of the Dvali-Gabadadze-Porrati scenario with a Gauss-Bonnet term in the bulk. We concentrate on the solution that generalises the normal Dvali-Gabadadze-Porrati branch. It is well known that this branch cannot describe the late-time acceleration of the universe even with the inclusion of a Gauss-Bonnet term. Here, we show that this branch in the presence of a Gauss-Bonnet curvature effect and an holographic dark energy with the Hubble scale as the infra-red cutoff can describe the late-time acceleration of the universe. It is worthwhile to stress that such an energy density component cannot do the same job on the normal  Dvali-Gabadadze-Porrati branch (without Gauss-Bonnet modifications) nor in a standard 4-dimensional relativistic model. The acceleration on the brane is also presented as being induced through an effective dark energy which corresponds to a balance between the holographic one and geometrical effects encoded through the Hubble parameter.
\end{abstract}


\date{\today}
\maketitle


\section{Introduction}

More than a decade ago, observations of supernovae type Ia led to a surprising discovery \cite{Perlmutter:1998np}, the universe has entered accelerated expansion recently; i.e. such observations could match the luminosity distance versus the redshift for a relativistic Friedmann-Lema\^{\i}re-Robertson-Walker (FLRW) universe if and only if a dark energy component or a cosmological constant is invoked on the budget of the universe. Later on, this conclusion was corroborated for example by measurements of  the
cosmic microwave background (CMB) \cite{Komatsu:2010fb} and the baryon acoustic oscillations (BAO) \cite{Cole:2005sx}. More recently,  gamma ray bursts (GRB), if not properly standard candles,  have also been  very useful in this regard \cite{izzo,Cardone:2010rr}. What is this component filling the universe? Is it a constant of nature? Is it evolving with time? Or is it a consequence of quantum gravity? We are far from giving an answer to these fundamental questions in this paper. We will rather follow a phenomenological approach inspired on a positive answer to the third question raised above \cite{footnoteA}, more precisely based on the holographic dark energy scenario \cite{Cohen:1998zx,Li:2004rb,Hsu:2004ri}.

Several years ago Bekenstein proposed a bound on the total entropy of a system of volume $L^3$; such a bound is proportional to  the area of the system,  $L^2$, rather than its volume \cite{Bekenstein:1973ur,pedro,pedro2}. Indeed, for such a  system the number of degrees of freedom would be finite and related to its area according to the holographic principle \cite{'tHooft:1993gx,Susskind:1994vu}. Consequently, a breakdown of the quantum field theory for large volumes would be expected \cite{Cohen:1998zx}: for an effective quantum field theory with an ultra-violet (UV) cutoff, $M_{\rm{UV}}$, the entropy of a system would scale as $L^3M_{\rm{UV}}^{3}$. To circumvent this breakdown, Cohen {\textit{ et al.}} proposed a relationship between UV and infrared (IR) cutoffs ensuring  the validity of an effective quantum field theory within this regime \cite{Cohen:1998zx}:
\begin{equation}
L^3M_{\rm{UV}}^4\lesssim LM_p^2.
\end{equation}
The largest scale allowed is the one saturating the previous inequality, for which we can define an energy density \cite{Li:2004rb,Hsu:2004ri}
\begin{equation}
\rho_{\mathrm{H}}=\frac{3c^2}{\kappa_4^2 L^2},
\label{rhoholographic}
\end{equation}
where for convenience the constant $3c^2$ has been introduced and $\kappa_4^2=8\pi G$. The holographic dark energy model is based on assuming the energy density $\rho_{\mathrm{H}}$  is responsible for the current speed up of the universe with $L$ being an appropriate cosmological length. Indeed, it was soon realised that by choosing $L$ as the  inverse of the current Hubble rate, $\rho_{\mathrm{H}}$ is of the order of the current dark energy density \cite{Li:2004rb}, even though the equation of state would not be the appropriate one as it  would not induce acceleration in a homogeneous and isotropic universe. Later on, $L$ was chosen to be the event horizon and it turns out that in this case the late-time acceleration of the universe is suitably described \cite{Hsu:2004ri}.

An alternative approach to explain the current acceleration of the universe is based on the brane-world scenario \cite{Maartens:2010ar}, which is inspired on string theory, a candidate to quantum theory. In this kind of scenario, our universe is a brane, i.e. a 4-dimensional (4d) hypersurface, embedded in the bulk; i.e. the higher dimensional space-time. The simplest and best studied case is the Dvali-Gabadadze-Porrati (DGP) model \cite{Dvali:2000hr}. Such a model contains two kinds of solutions \cite{Deffayet:2000uy}: the self-accelerating branch, which suffers from some problems, and the normal branch. Even though the normal branch is \textit{healthy} it cannot describe the current acceleration of the universe unless a dark energy component is invoked \cite{BouhmadiLopez:2007ts} or the gravitational action is modified \cite{BouhmadiLopez:2010pp}. 

Although extra-dimensions have not been detected experimentally, they have a theoretical basis on string theory, which might or might not be the ultimate theory of nature. We are far to address this issue on the present paper. Brane-world models are a simple arena where to test ``string inspired'' models through cosmology, for example. The induced gravity term (or a 4d Einstein-Hilbert term in the classical theory) is expected to arise on the brane action as a result of the loop-corrections from matter fields \cite{Dvali:2000hr}. On the other hand, a Gauss-Bonnet curvature term is as well expected to arise in the bulk, motivated from string theory and also  from a mathematical/physical point of view as it gives raise to equations of motion of second order in the metric. Both effects were first analysed in reference \cite{Kofinas:2003rz} and latter on in references \cite{richard,BouhmadiLopez:2008nf}, where it was shown that they can be quite useful to describe the current acceleration of the universe, for example they can  mimic a phantom like behaviour without phantom matter,  and appease some of the singularities that can take place in the universe like the big bang  or the big rip. Finally, the holographic dark energy is a phenomenological approach to implement the holographic principle in a cosmological setting. The main aim of this paper is to show that even though an holographic energy density with the Hubble rate as the infra-red cutoff is not suitable to describe the late-time acceleration of a 4d relativistic model nor a brane-world model with induced gravity (on the normal branch), it is suitable to describe the late-time acceleration of a brane world-model with an induced gravity term on the brane and a Gauss-Bonnet effect in the bulk.

An holographic dark energy model within the DGP scenario was analysed in \cite{Wu:2007tp} (see also \cite{saridakis}), where it was shown that the normal branch filled with an holographic energy density with cutoff scale $L=H^{-1}$ cannot explain the current acceleration of the universe. In this paper, we will show that by invoking a Gauss-Bonnet term in the bulk the situation can be improved and the brane undergoes a speed up at late-time. 

The outline of the paper is as follows. In Sect.~II,  we present the holographic brane-world model we analyse, we impose some constraints on the parameters of the model and deduce the maximally symmetric solutions of the model. In Sect.~III, we analyse the modified Friedmann equation of the model. By solving this cubic equation analytically, we are able to obtain the different evolutions the brane can undergo. In Sect.~III, we pick up the most suitable brane to describe the current acceleration. 
We present as well a description of the effective dark energy responsible for the late-time acceleration of the brane, in Sect.~IV. Finally, in Sect.~V, we summarise and present our conclusions.

\section{The setup and parameter constraints}

We consider a 5-d brane-world model, where the bulk contains a Gauss-Bonnet curvature term on top of the usual Hilbert-Einstein action, while the brane contains only an induced gravity term on its action. Then the modified Friedmann equation reads \cite{Kofinas:2003rz,richard}
\begin{equation}
H^2=\frac{\kappa_4^2}{3}\rho+\frac{\epsilon}{r_c}\left(1+\frac{8\alpha}{3}H^2\right)H,
\label{friedmann1}
\end{equation}
$r_c$ and $\alpha$ correspond to the cross-over scale  and the Gauss-Bonnet parameter, respectively. On the other hand, $\epsilon=\pm 1$. For $\epsilon=-1$ and $\alpha=0$ we recover the normal DGP branch while for $\epsilon=1$ and $\alpha=0$ we obtain the self-accelerating DGP branch \cite{Deffayet:2000uy}. As we are dealing with the late-time acceleration of the universe, the matter content of the brane can be described through a cold dark matter component (CDM) with energy density $\rho_\mathrm{m}$ and a dark energy component given by an holographic energy density, $\rho_{\mathrm{H}}=3c^2/{\kappa_4^2 L^2}$ ,  (cf.~\cite{Cohen:1998zx,Li:2004rb}).
As already mentioned in the introduction, $\rho_{\mathrm{H}}$ is related to the UV cutoff, while $L$ is related to the IR cutoff. The parameter $L$ was first identified with the inverse of the Hubble rate, even though it was shown later on in \cite{Hsu:2004ri} that such a choice could not describe the current acceleration of the universe in a 4-d relativistic model and other proposals were put forward \cite{Li:2004rb}. From now on we will only consider the brane with $\epsilon =-1$ because the other branch ($\epsilon =+1$) is self-accelerating on the absence of any kind of dark energy \cite{richard}.

In this paper, we will show that by considering a brane-world model with an induced gravity term on the brane and a Gauss-Bonnet term in the bulk, the brane undergoes a consistent acceleration induced by an holographic dark energy with $L=H^{-1}$. Notice that this cannot take place on the normal DGP branch \cite{Wu:2007tp}. The Friedmann equation (\ref{friedmann1}) can be written more conveniently as 
\begin{equation}
(1-c^2)E^2(z)=\Omega_m(1+z)^3-2\sqrt{\Omega_{r_c}}(1+\Omega_\alpha E^2(z))E(z),
\label{friedmann2}
\end{equation}
where $E(z)=H/H_0$ and
\begin{eqnarray}
\Omega_m=\frac{\kappa_4^2 \rho_{m_0}}{3H_0^2},\,\,\,\, 
\Omega_{r_c}=\frac{1}{4r_c^2H_0^2},\,\,\,\, \Omega_{\alpha}=\frac{8}{3}\alpha H_0^2\,,\\ \nonumber
\label{defomega}
\end{eqnarray}
are the usual convenient dimensionless parameters (we will follow the notation of  \cite{BouhmadiLopez:2008nf}). The energy density of matter is conserved on the brane as we are assuming a maximally symmetric bulk and we have not assumed any sort of interaction between matter and the holographic component on the brane. We assume that the holographic dark energy stated in equation (\ref{rhoholographic}) remains valid on brane-world models like the one we have considered here. A similar approach was followed on \cite{Wu:2007tp}. Our approach is nevertheless consistent with the Bianchi identity, which is satisfied, as the total energy density is conserved on the brane (see Eq.~(4.5)), more precisely, the matter energy density as well as the effective energy density are conserved (cf. Eq.~(4.3)). In addition, the holographic energy we consider do not violate the null energy condition (or the dominant energy condition) because the energy density is proportional to the square of the Hubble rate which is a decreasing function (as a function of time) for the set of physical solutions we analyse in section IV. We, therefore, do not have the kind of instabilities present on phantom models.

In absence of matter, i.e. $\Omega_m=0$, we obtain two kind of maximally symmetric solutions: a flat solution; i.e. $E=0$, and two de Sitter  solutions with dimensionless Hubble rates:
\begin{equation}
E_\pm=\frac{c^2-1\pm\sqrt{(c^2-1)^2-16\Omega_{r_c}\Omega_\alpha}}{4\Omega_\alpha\sqrt{\Omega_{r_c}}}.
\label{solEpm}
\end{equation}
Notice that the de Sitter solutions exist if and only if $16\Omega_{r_c}\Omega_\alpha<(c^2-1)^2$ \cite{footnote1}.

The cosmological parameters of the model are constrained by evaluating the Friedmann equation (\ref{friedmann2}) at $z=0$
\begin{equation}
\Omega_m-2\sqrt{\Omega_{r_c}}(1+\Omega_\alpha) +c^2=1.
\label{constraintom}
\end{equation}
The deceleration parameter, $q=-a\ddot{a}/{\dot a}^2$, can be obtained and in particular its present value reads
\begin{equation}
q_0=-\left(1-\frac32 \frac{\Omega_m}{(1-c^2)+\sqrt{\Omega_{r_c}}(1+3\Omega_\alpha)}\right).
\label{q0}
\end{equation}
By combining Eqs.~(\ref{constraintom}) and (\ref{q0}), we obtain
\begin{equation}
q_0=\frac12 \frac{\Omega_m+2\sqrt{\Omega_{r_c}}(1-\Omega_\alpha)}{\Omega_m-\sqrt{\Omega_{r_c}}(1-\Omega_\alpha)}.
\label{q02}
\end{equation}
The universe is currently accelerating, therefore $q_0<0$. Bearing in mind that the current value of $q_0$ is roughly $q_0\sim - 0.7$,  we conclude that
\begin{equation}
0<4\sqrt{\Omega_{r_c}}<c^2-1.
\label{conditionc}
\end{equation}
Therefore, the parameter $c$ must satisfy $1<c^2$ but still not too large. This condition will simplify considerably our analysis of the modified Friedmann equation (\ref{friedmann2}). 

It is worthwhile to look as well at the Raychaudhuri equation which reads
\begin{equation}
\frac{\dot E}{H_0}=-\frac32\frac{\Omega_m (1+z)^3 E(z)}{(1-c^2)E(z)+\sqrt{\Omega_{r_c}}(1+3\Omega_\alpha E^2(z))},
\label{Raychaurudi}
\end{equation}
where a dot stands for the derivative with respect to the cosmic time of the brane. As $1<c^2$, it turns out that $\dot E$ might diverge at the finite dimensionless Hubble rates
\begin{equation}
E_{s_{1,2}}=\frac{c^2-1\pm\sqrt{(c^2-1)^2-12\Omega_{r_c }\Omega_\alpha}}{6\Omega_\alpha\sqrt{\Omega_{r_c}}}\;,
\label{hubblesudden}
\end{equation}
where $E_{s_{1}}$ and $E_{s_{2}}$ stand for the expression with a minus and a plus in front of the square root, respectively. The blowing up of $\dot H$ may indicate the presence of sudden singularities \cite{Nojiri:2005sx,Cattoen:2005dx}. It is well known that in some extra-dimensional models such a behaviour can show up even for a finite pressure \cite{Shtanov:2002ek,richard,BouhmadiLopez:2008nf,BouhmadiLopez:2010vi}. It can be shown that if the dimensionless Hubble rates  $E_{s_{1}}$ and $E_{s_{2}}$ are reached they take place at the (dimensionless) energy densities $\bar\rho_1$ and $\bar\rho_2$, respectively, defined in Eqs.~(\ref{defbarrho1})-(\ref{defbarrho2}) (see as well Eq.~(\ref{dimensionlessrho}) and Fig.~\ref{prho1rho2}). It will be clear on the next section when this can be the case.

\section{The HDGP-GB normal branch}

\begin{figure}[t]
\begin{center}
\includegraphics[width=0.7\columnwidth]{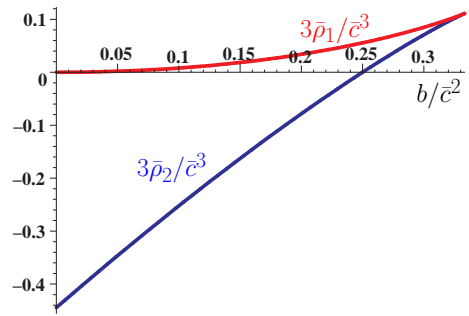}
\end{center}
\caption{Plot of the (rescaled) dimensionless energy densities $\bar\rho_1$ and $\bar\rho_2$ against $b/\bar{c}^2$.}
\label{prho1rho2}
\end{figure}

In order to continue the analysis of this model we need to solve the cubic Friedmann equation (\ref{friedmann2}) which we tackle next.  Concerning this aim, it is convenient to introduce the dimensionless variables:
\begin{eqnarray}
\bar H&=&\frac83 \frac{\alpha}{r_c}H = 2\Omega_\alpha\sqrt{\Omega_{r_c}}E(z), \\
\bar\rho&=&\frac{32}{27}\frac{\kappa_5^2\alpha^2}{r_c^3}\rho_m=4\Omega_{r_c}\Omega_\alpha^2\Omega_m(1+z)^3,\label{dimensionlessrho}\\
 b&=&\frac83\frac{\alpha}{r_c^2}=4\Omega_\alpha\Omega_{r_c}, \label{defb}\\
\bar c&=& c^2-1.\label{deftildec}
\label{dimensionlessq}
\end{eqnarray}
The Friedmann equation can be rewritten as follows
\begin{equation}
{\bar H}^3- \bar c {\bar H}^2+b\bar H-\bar\rho=0.
\label{Friedmannnb}
\end{equation}
Notice that $\bar c$ has to be positive in order for the brane to be accelerating at present, as was shown on the previous section. The number of real roots of the previous equation is determined by the sign of the discriminant function
${\mathcal{N}}$
defined as \cite{Abramowitz}
\begin{equation}
{\mathcal{N}}= Q^3+R^2,
\label{NRR1}
\end{equation}
where $Q$ and $R$ read
\begin{equation}
Q=\frac{{\bar c}^2}{3}\left(\frac{b}{{\bar c}^2}-\frac13\right),\quad R=-\frac{{\bar c}^3}{3}\left( \frac{b}{{2\bar c}^2}-\frac32\frac{\bar\rho}{{\bar c}^3} -\frac{1}{9}\right).
\label{QR}
\end{equation}
It is
helpful to rewrite ${\mathcal{N}}$  as follows
\begin{equation}
{\mathcal{N}} =\frac14(\bar\rho-\bar\rho_1)(\bar\rho-\bar\rho_2),
\label{NRR2}
\end{equation}
where
\begin{eqnarray}
\bar\rho_1&=&\frac{{\bar c}^3}{3}\left\{\frac{b}{{\bar c}^2}-\frac29\left[1-\sqrt{\left(1-3\frac{b}{{\bar c}^2}\right)^3}\right]\right\},\label{defbarrho1} \\
\bar\rho_2&=&\frac{{\bar c}^3}{3}\left\{\frac{b}{{\bar c}^2}-\frac29\left[1+\sqrt{\left(1-3\frac{b}{{\bar c}^2}\right)^3}\right]\right\},
\label{defbarrho2}
\end{eqnarray}
for the analysis of the number of physical solutions of the modified Friedmann equation (\ref{Friedmannnb}).
If ${\mathcal{N}}$  is positive then there is a unique real solution.
On the other hand, if ${\mathcal{N}}$  is negative there are 3 real solutions. Finally,
if ${\mathcal{N}}$  vanishes, all roots are real and at least two are equal. For a given $\bar\rho$, the ratio $b/{\bar c}^2$ determines completely the sign of ${\mathcal{N}}$ as a consequence of the positiveness of $\bar c $ [cf. Eqs.~(\ref{conditionc}) and (\ref{deftildec}) and Fig.~\ref{prho1rho2}].

\subsection{Case 1: $0<b/{\bar c}^2<1/4$}\label{physical}

In this case, $0<\bar\rho_1$, $\bar\rho_2<0$ and the solutions can be split in three subsets:

\begin{figure}[t]
\begin{center}
\includegraphics[width=0.7\columnwidth]{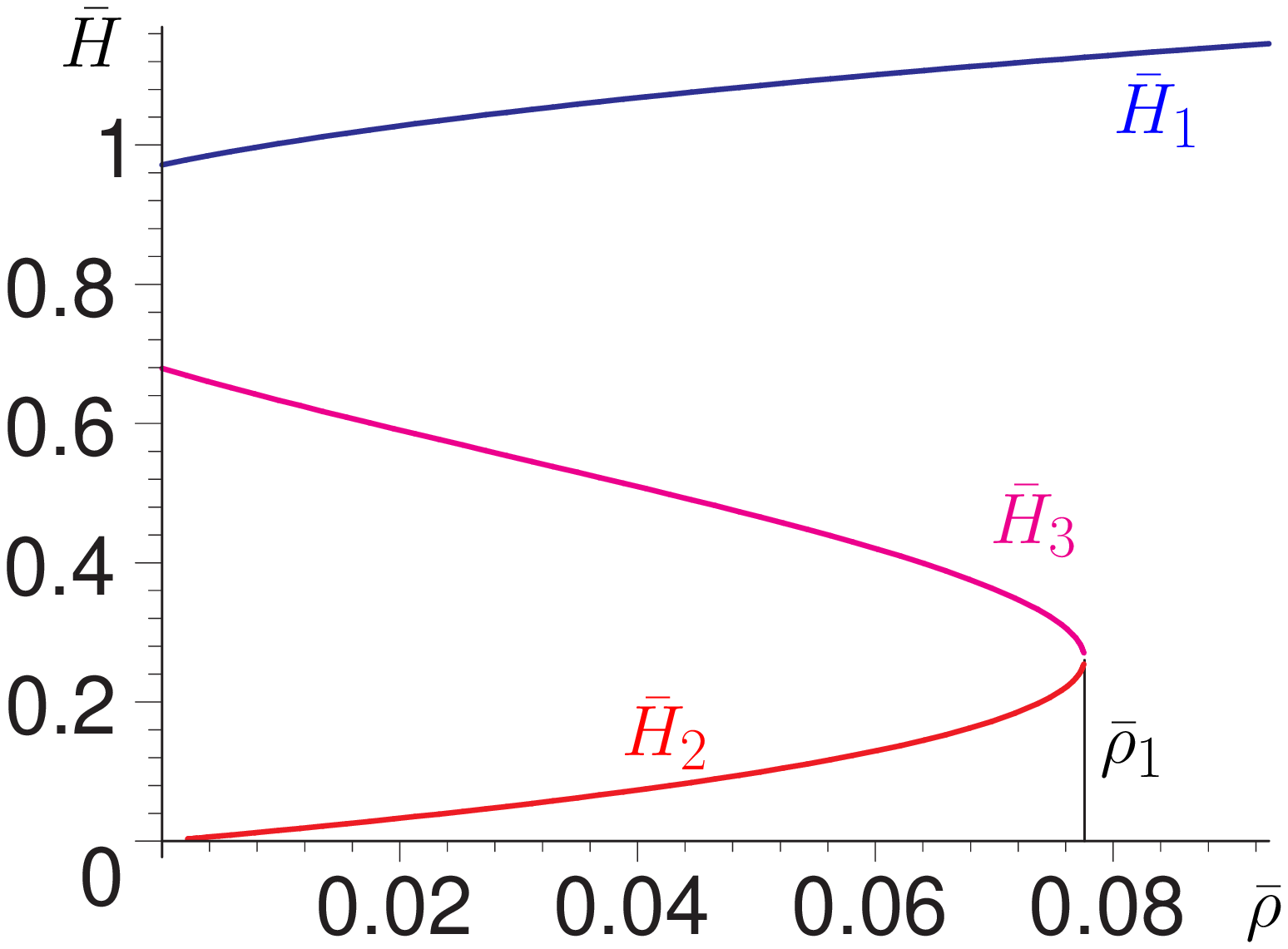}
\end{center}
\caption{Plot of the dimensionless Hubble rates $\bar H_1$, $\bar H_2$ and $\bar H_3$ against the dimensionless energy density $\bar\rho$ for \mbox{$0<b/{\bar c}^2<1/4$}.}
\label{pless14}
\end{figure}

\subsubsection{$0\leq \bar\rho<\bar\rho_1$}

We have three positive solutions for the dimensionless Hubble rate $\bar H$ as long as $\bar\rho<\bar\rho_1$, which read
\begin{eqnarray}
{\bar H}_1&=& \frac23\left[\sqrt{{\bar c}^2-3b}\cos\left(\frac{\theta-\pi}{3}\right) +\frac{\bar c}{2}\right], \label{H11}\\
{\bar H}_2&=& -\frac23\left[\sqrt{{\bar c}^2-3b}\cos\left(\frac{\theta}{3}\right) -\frac{\bar c}{2}\right], \label{H21}\\
{\bar H}_3&=& \frac23\left[\sqrt{{\bar c}^2-3b}\cos\left(\frac{\theta+\pi}{3}\right) +\frac{\bar c}{2}\right], \label{H31}
\end{eqnarray}
where
\begin{equation}
 \cos(\theta)=\frac{R}{\sqrt{-Q^3}},\quad \sin(\theta)=\sqrt{\frac{Q^3+R^2}{Q^3}},\quad 0\leq\theta_0\leq\theta<\pi.
\end{equation}
The angle $\theta$ is an increasing function of the dimensionless energy density $\bar\rho$. Indeed, at $\bar\rho=0$ or in the far future (redshift $z=-1$), $\theta=\theta_0$, where
\begin{equation}
\cos\left(\frac{\theta_0}{3}\right)=\frac12 \frac{\bar c}{\sqrt{\bar c^2-3b}}, 
\quad \sin\left(\frac{\theta_0}{3}\right)=\frac{\sqrt{3}}{2}\sqrt{\frac{\bar c^2-4b}{\bar c^2-3b}}.
\end{equation}
Therefore, $\bar H_2|_{\theta_0}=0$ and 
\begin{equation}
\bar H_1|_{\theta_0}=\frac12\left[\bar c +\sqrt{{\bar c}^2-4b}\right], \quad \bar H_3|_{\theta_0}=\frac12\left[\bar c -\sqrt{{\bar c}^2-4b}\right].
\label{H1H3theta0}
\end{equation}
These solutions are simply the maximally symmetric solutions we obtained in section II, where the Minkowski solution is  $\bar H_2|_{\theta_0}$, while the de Sitter solution $E_+$ and $E_-$ (given in Eq.~(\ref{solEpm})) correspond to $\bar H_1|_{\theta_0}$ and $\bar H_3|_{\theta_0}$, respectively.

It can be shown that $0\leq\bar H_2<\bar H_3<\bar H_1$ (cf. Fig.~\ref{pless14}). Then when $\bar\rho$ approaches $\bar\rho_1$, $\theta\rightarrow \pi$ where $\bar H_2$ overlaps with $\bar H_3$.

\subsubsection{$\bar\rho_1=\bar\rho$}

This is a very particular situation where $\mathcal{N}$ vanishes and consequently there are only two different solutions:
\begin{eqnarray}
\bar H_1&=&\frac13\left[2\sqrt{\bar c^2-3b}+\bar c\right], \label{H12}\\
\bar H_2&=&\bar H_3= \frac13\left[-\sqrt{\bar c^2-3b}+\bar c\right]\label{H22},
\end{eqnarray}
which  are the analytical continuation of the solutions presented in the previous subsection as Fig.~\ref{pless14} shows clearly.
The solution $\bar H_2$ or $\bar H_3$ corresponds to the Hubble rate $E_{s_2}$ given in Eq.(\ref{hubblesudden}). At the finite positive energy density  $\bar\rho_1$ the cosmic derivative of the Hubble rate blows up even though the pressure is zero; consequently when $\bar H_2= \bar H_3$ there is a sudden singularity \cite{Nojiri:2005sx}.

For completeness, we will discuss what happens at the other Hubble rate $E_{s_1}$ (see Eq.~(\ref{hubblesudden})) where again $\dot H$ blows up. If ${\it 0<b/\bar c^2<1/4}$, that value of the Hubble rate is reached for the  negative energy density $\bar\rho_2$. This happens when the solutions $\bar H_1$ and $\bar H_3$, plotted in Fig.~\ref{pless14}, become equal which takes place for negative energies on the left hand side continuation of figure \ref{pless14}. This case is unphysical on the setup  ${\it 0<b/\bar c^2<1/4}$ and we will discard it.

\subsubsection{$\bar\rho_1<\bar\rho$}

As figure \ref{prho1rho2} shows, there is a unique real solution for $\bar\rho_1<\bar\rho$ (The discriminant $\mathcal{N}$ is positive)
\begin{equation}
{\bar H}_1=\frac13\left[2\sqrt{{\bar c}^2-3b}\cosh\left(\frac{\eta}{3}\right)+\bar c\right],
\label{H13}
\end{equation}
where 
\begin{equation}
\cosh(\eta)= -\frac{R}{\sqrt{-Q^3}}, \quad \sinh(\eta)= \sqrt{\frac{Q^3+R^2}{-Q^3}},\,\,\,\, 0<\eta.
\end{equation}
The parameter $\eta$ is an increasing function of the dimensionless energy density $\bar\rho$. In particular, when $\eta\rightarrow 0$, $\bar\rho \rightarrow \bar\rho_1$. At that point the solution (\ref{H13}) connects with the solution $\bar{H}_1$ given in Eq.~(\ref{H12}).

In summary for $b/{\bar c}^2<1/4$, there are three kinds of solutions: (i) the solution $\bar H_1$ which starts in the far past, through a big bang singularity, expanding and is asymptotically de Sitter in the future, (ii)  the solution $\bar H_2$ which starts through a sudden singularity and keeps expanding until it reaches a Minkowski state and (iii)  the solution $\bar H_3$ which starts also through a sudden singularity then it follows  a super-inflationary expansion; i.e. $0<\dot{{\bar H}}_3$, until it becomes asymptotically de Sitter in the future.

\subsection{Case 2: $b/{\bar c}^2=1/4$}

\begin{figure}[t]
\begin{center}
\includegraphics[width=0.7\columnwidth]{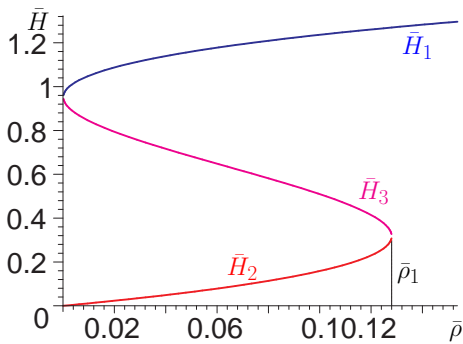}
\end{center}
\caption{Plot of the dimensionless Hubble rates $\bar H_1$, $\bar H_2$ and $\bar H_3$ against the dimensionless energy density $\bar\rho$ for $b/{\bar c}^2=1/4$.}
\label{pequal14}
\end{figure}

In this case $0<\bar\rho_1$ and $\bar\rho_2=0$, therefore there are three kinds of solutions which can be split in three subsets like for the case $b/{\bar c}^2<1/4$. In fact, the three solutions can be denoted $\bar H_1$, $\bar H_2$, and $\bar H_3$ and can be written using the same analytical expressions given in Eqs.~(\ref{H11})-(\ref{H31}), (\ref{H12}), (\ref{H22}) and (\ref{H13}) with  the constraint $b/{\bar c}^2=1/4$, even though they are physically different. An example of such set of solutions is given in Fig.~\ref{pequal14}.

The brane with Hubble rate $\bar H_1$ starts with a big bang singularity and expands until it hits a sudden singularity in the future when $\bar\rho=\bar\rho_2=0$. The solution with Hubble rate $\bar H_2$ starts its cosmological evolution from a sudden singularity and expands until becoming asymptotically Minkowski in the future. Finally, the brane with Hubble rate $\bar H_3$ starts at a sudden singularity with $\bar\rho=\bar\rho_1$ and expands in a super-inflationary way until it hits a sudden singularity at $\bar\rho=\bar\rho_2=0$.

It is worthwhile to clarify why the solutions  $\bar H_1$ and $\bar H_3$ finish their classical evolution at a sudden singularity when $\bar\rho=\bar\rho_2=0$ rather than in a de Sitter expansion as happens for $b/{\bar c}^2<1/4$. The reason is simply that at that energy density, the Hubble rates $\bar H_1|_{\theta_0}$ and $\bar H_3|_{\theta_0}$ (see Eq.~(\ref{H1H3theta0}) and Ref.~\cite{footnote2}) are equal and coincide with the Hubble rate $2\Omega_\alpha\sqrt{\Omega_{r_c}}E_{s_1}$ where $\dot H$ diverges.

\begin{figure}[t]
\begin{center}
\includegraphics[width=0.7\columnwidth]{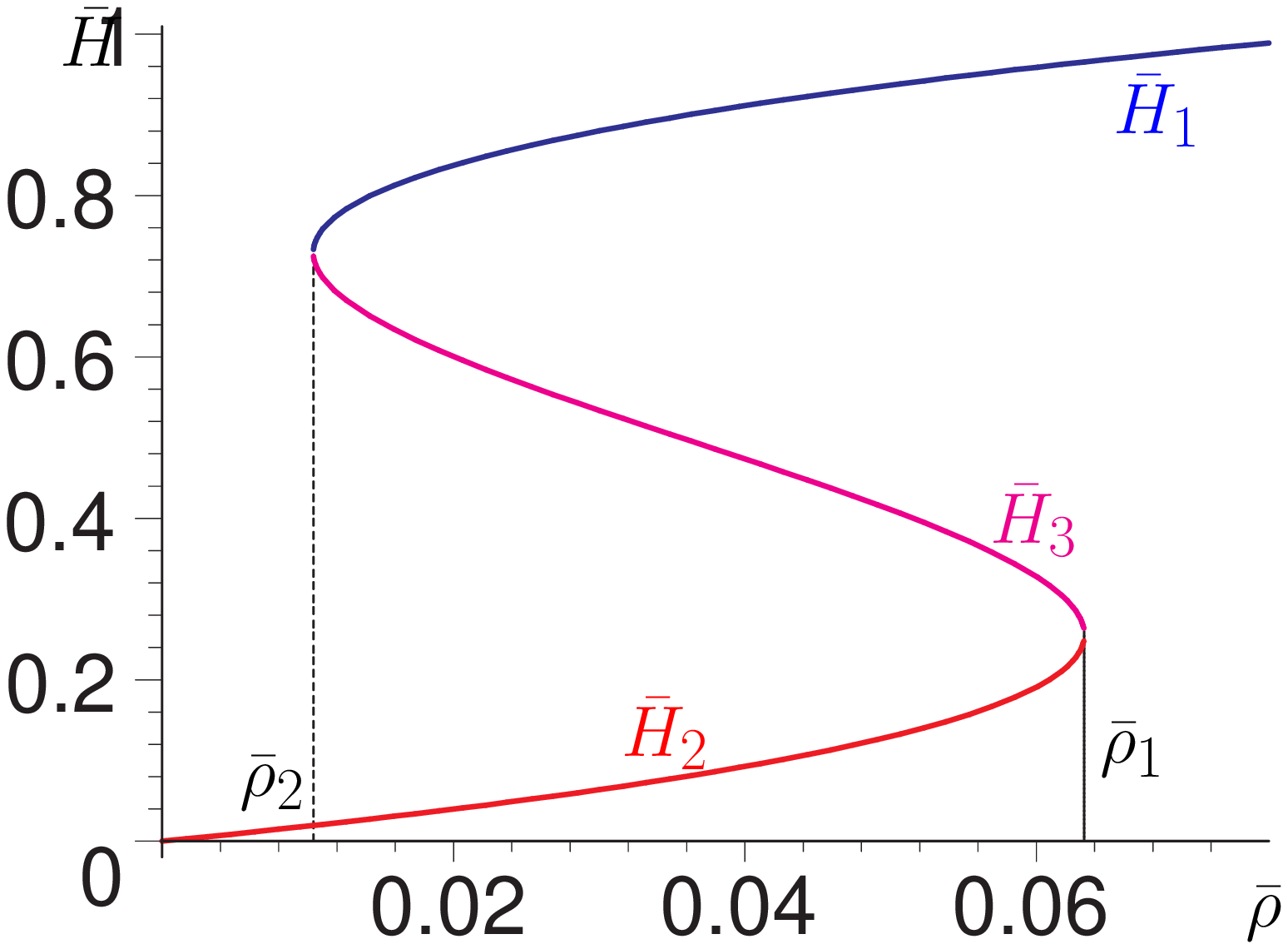}
\end{center}
\caption{Plot of the dimensionless Hubble rates $\bar H_1$, $\bar H_2$ and $\bar H_3$ against the dimensionless energy density $\bar\rho$ for \mbox{$1/4<b/{\bar c}^2<1/3$}.}
\label{pbetween1413}
\end{figure}

\subsection{Case 3: $1/4<b/{\bar c}^2<1/3$}

In this case $\bar\rho_1$ and  $\bar\rho_2$ are positive as Fig.~\ref{prho1rho2} shows. This case is slightly more cumbersome but can be worked out in analogy with the previous cases. There are three kinds of solutions (cf. Fig~\ref{pbetween1413}):

The solution $\bar H_1$ starts in the far future at a big bang singularity and expands until it hits a sudden singularity at $\bar\rho_2$. This solution is described by the expressions (\ref{H11}), (\ref{H12}) and  (\ref{H13}) depending on the value of the energy density \cite{footnote3}.

The solution $\bar H_2$ starts at a sudden singularity with a finite energy density $\bar\rho_1$ and finite pressure (indeed the pressure vanishes for CDM) although $\dot H$ diverges. Then it expands until it becomes asymptotically de Sitter in the future. For $0<\bar\rho_2\leq\bar\rho\leq\bar\rho_1$ the solution is described by the equations (\ref{H21}) and (\ref{H22}) \cite{footnote3}. For $\bar\rho<\bar\rho_2$ the solution reads
\begin{equation}
\bar H_2=-\frac13\left[2\sqrt{{\bar c}^2-3b}\sinh\left(\frac{\gamma}{3}\right)-\bar c\right]
\end{equation}
where
\begin{equation}
\cosh(\gamma)=\frac{R}{\sqrt{-Q^3}}, \quad \sinh(\gamma)=\sqrt{\frac{Q^3+R^2}{-Q^3}},\,\, 0<\gamma<\gamma_0.
\end{equation}
The variable $\gamma$ is a decreasing function of $\bar\rho$. When $\bar\rho$ is vanishing $\gamma\rightarrow \gamma_0$. On the other hand, when $\bar\rho\rightarrow\bar\rho_2$ the variable $\gamma$ fulfils $\gamma\rightarrow 0^+$.

Finally, the solution $\bar H_3$ starts at a sudden singularity where $\bar\rho=\bar\rho_1$ and ends at another sudden singularity where $\bar\rho=\bar\rho_2$. The whole expansion of the brane is super-inflationary on this case; i.e. the Hubble rate grows as the brane expands, and it is described by Eqs.~(\ref{H31}) and (\ref{H22}) \cite{footnote3}.

\subsection{Case 4: $1/3\leq b/{\bar c}^2$}

This is the simplest case to analyse as $\bar\rho_1$ and $\bar\rho_2$ are complex and conjugate numbers and therefore $\mathcal{N}$ is positive, implying therefore a unique real solution which reads
\begin{equation}
\bar H_1=-\frac13\left[2\sqrt{3b-\bar c^2}\sinh\left(\frac{\delta}{3}\right)-\bar c\right],
\end{equation}
where 
\begin{equation}
\cosh(\delta)=\sqrt{\frac{Q^3+R^2}{Q^3}}, \quad \sinh(\delta)=\frac{R}{\sqrt{Q^3}}, \,\,\,\, \delta\leq\delta_0.
\end{equation}
When $\bar\rho\rightarrow 0$, $\delta\rightarrow\delta_0$. On the other hand, for very large values of the energy density, $\delta\rightarrow -\infty$. The brane starts with a big bang singularity. It keeps expanding until it becomes asymptotically Minkowski.

\section{Effective dark energy}
\label{Effective dark energy}

\begin{figure}[t]
\begin{center}
\includegraphics[width=0.7\columnwidth]{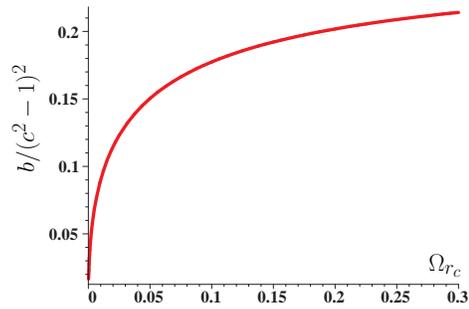}
\end{center}
\caption{Plot of the  parameter $b/\bar c^2=b/(c^2-1)^2$ against  $\Omega_{r_c}$. The cosmological parameters are assumed to be $\Omega_m=0.27$ and $q_0=-0.7$. The rest of the parameters are fixed as explained on Sec.~\ref{Effective dark energy}.}
\label{pbcc}
\end{figure}

We showed on the previous section that there are different solutions which are characterised by the parameter $b/(c^2-1)^2$. The question now is which one is the most suitable to describe the current acceleration of the universe. The ratio $b/(c^2-1)^2$ is determined by the  content of the Universe. In this regard we know that $\Omega_m=0.27$ and $q_0 \sim -0.7$, where we have used the latest WMAP7 data \cite{Komatsu:2010fb} and considered that our model does not  deviate too much from the $\Lambda$CDM scenario at the present time to estimate the current deceleration parameter. Therefore, for a given $\Omega_{r_c}$, we can easily determine $\Omega_\alpha$ through Eq.~(\ref{q02}) and subsequently the holographic parameter $c$ using the constraint (\ref{constraintom}). In summary, following this procedure  we can determine the range of suitable values for  $b/(c^2-1)^2$. We show our results on Fig.~\ref{pbcc}. As can be deduced from this plot $0<b/(c^2-1)^2<1/4$. Hence, for this range of parameters,  there are only three possible solutions analysed on subsection \ref{physical} and plotted in Fig.~\ref{pless14}. The solutions $\bar H_2$ and  $\bar H_3$ cannot describe the history of the universe because the dimensionless energy density $\bar\rho_1$ takes place at a very low redshift or even in the future \cite{footnote4}. In addition, The brane with Hubble rate $\bar H_3$ is self-accelerating along its whole expansion which is not physical. Therefore, the only suitable solution is $\bar H_1$ (cf. Eqs.~(\ref{H11}), (\ref{H12}) and (\ref{H13})). \textit{Can this solution describe the current acceleration of the universe and at the same time be consistent with the earlier evolution of the universe, like the Big Bang Nucleosynthesis (BBN)  epoch?} we next reply to these questions.

It is helpful to rewrite the Friedmann equation (\ref{friedmann2}) by
defining a corresponding effective energy density $\rho_{\rm{eff}}$ and
an effective equation of state with parameter $w_{\rm{eff}}$. More
precisely, the effective description is inspired by writing
down the modified Friedmann equation of the brane as
the usual relativistic Friedmann equation so that
\begin{equation}
H^2=\frac{\kappa_4^2}{3}(\rho_m+\rho_{\rm{eff}}),
\label{friedmanneff}
\end{equation}
i.e. to map the brane evolution in Eq.~(\ref{friedmann2}) to an equivalent 4-d general relativistic model with
Friedmann equation (\ref{friedmanneff}). In the previous equation the effective energy density $\rho_{\rm{eff}}$ reads
\begin{equation}
\rho_{\rm{eff}}=\frac{3H_0^2}{\kappa_4^2}\left\{c^2E^2(z)-2\sqrt{\Omega_{r_c}}\left[1+\Omega_{\alpha} E^2(z)\right]E(z)\right\}.
\label{rhoeff}
\end{equation}
This effective energy density corresponds to a balance between the holographic energy density and geometrical effects encoded on the Hubble parameter. The dependence of $\rho_{\rm{eff}}$ on the redshift is known analytically by means of the solutions $\bar H_1$ of the cubic Friedmann equation presented in the previous section.
\begin{figure}[t]
\begin{center}
\includegraphics[width=0.7\columnwidth]{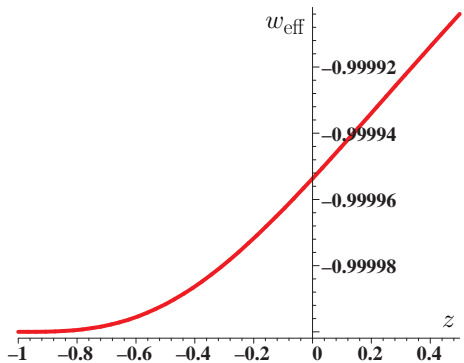}
\end{center}
\caption{Plot of  the parameter $w_{\rm{eff}}$ against the redshift $z$. The cosmological parameters are assumed to be $\Omega_m=0.27$, $\Omega_{r_c}=0.25$ and $q_0=-0.7$}
\label{pweff}
\end{figure}
We can define an effective equation of state or parameter $w_{\rm{eff}}$ associated to
the effective energy density as
\begin{equation}
\dot\rho_{\rm{eff}}+3H(1+w_{\rm{eff}})\rho_{\rm{eff}}=0.
\end{equation}
The effective equation of state is defined in analogy with the standard relativistic case. By using Eq.~(\ref{rhoeff}), we obtain
\begin{eqnarray}
1+w_{\rm{eff}}&=&\frac{c^2E(z)-\sqrt{\Omega_{r_c}}(1+3\Omega_{\alpha}E^2(z))}{c^2E^2(z)-2\sqrt{\Omega_{r_c}}(1+\Omega_{\alpha}E^2(z))E(z)}\times\nonumber\\
&\;&\frac{\Omega_m(1+z)^3}{E(z)(1-c^2)+\sqrt{\Omega_{r_c}}(1+3\Omega_{\alpha}E^2(z))}.
\label{weff}
\end{eqnarray}
In Fig.~\ref{pweff}, we can see an example of the behaviour of $w_{\rm{eff}}$ for the current cosmological values. At very late-time $w_{\rm{eff}}$ approaches $-1$ as can be clearly proved from Eq.~(\ref{weff}) bearing in mind that the solution $\bar{H}_1$ is asymptotically de Sitter.

\begin{figure}[t]
\begin{center}
\includegraphics[width=0.7\columnwidth]{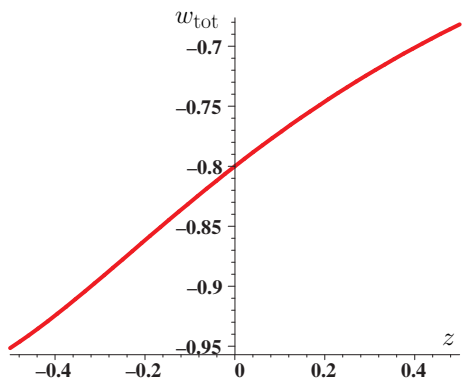}
\end{center}
\caption{Plot of  the parameter $w_{\rm{tot}}$ against the redshift $z$. The cosmological parameters are assumed to be $\Omega_m=0.27$, $\Omega_{r_c}=0.25$ and $q_0=-0.7$}
\label{pwl}
\end{figure}

Similarly, we can define the parameter $w_{\rm{tot}}$ 
\begin{eqnarray}\label{wtot}
w_{\rm{tot}}&=&-1-\frac23\frac{\dot H}{H^2},\\
&=&-1+\frac{\Omega_m(1+z)^3}{E^2(z)(1-c^2)+\sqrt{\Omega_{r_c}}(1+3\Omega_{\alpha}E^2(z))E(z)},\nonumber
\end{eqnarray}
which characterises the total equation of state of the brane and is different from $w_{\rm{eff}}$. Here also, $w_{\rm{tot}}$ approaches $-1$ at very late-time as can be easily deduced from Eq.~(\ref{wtot}) (see also Fig.~\ref{pwl}).

\begin{figure}[t]
\begin{center}
\includegraphics[width=0.7\columnwidth]{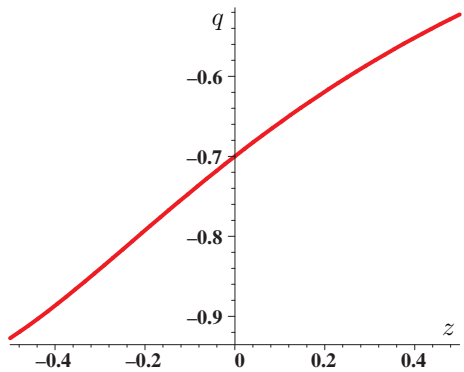}
\end{center}
\caption{Plot of the deceleration parameter $q$ against the redshift $z$ with the same assumption used in Figs.~\ref{pweff} and \ref{pwl}.}
\label{pq}
\end{figure}

In summary, what we have shown is that the solution $\bar H_1$ can accommodate the late-time acceleration of the brane (cf. Eqs.~(\ref{H11}), (\ref{H12}) and (\ref{H13}), and Fig.~\ref{pq}). Indeed, even though an holographic energy density in a 4-d relativistic model or in the normal DGP branch is not able to describe the current acceleration of the universe if the infra-red cutoff corresponds to the inverse of the Hubble rate \cite{Hsu:2004ri,Wu:2007tp}, it is no longer the case in the model we are analysing as Fig.~\ref{pq} shows clearly. 

Can the solution $\bar{H}_1$ be consistent with BBN as well? The expansion of the brane will be consistent with BBN as long as at that period of the cosmic expansion satisfies
\begin{equation}
 3 H^2\simeq \kappa_4^2\rho,
\end{equation}
or equivalently (see Eq.~(\ref{friedmann2})),
\begin{equation}
E^2(z)\gg |c^2E^2(z)-2\sqrt{\Omega_{r_c}}(1+\Omega_\alpha E^2(z))E(z)|.
\label{condBBN1}
\end{equation}
It turns out that at very high redshift this condition cannot be fullfilled because the cubic term on the Hubble rate in the modified equation dominates over the quadratic ones.

In summary, the answer to the question: ``\textit{Can the solution $\bar H_1$ (cf. Eqs.~(\ref{H11}), (\ref{H12}) and (\ref{H13})) (1) describe the current acceleration of the universe and (2) at the same time be consistent with the standard Big Bang Nucleosynthesis (BBN)  epoch?}'' is yes for the first point and no for the second point. While this model is suitable to describe the current epoch, it needs (may be) high energy corrections to the Gauss-Bonnet term to be suitable as well during the BBN.

\section{Conclusions}

We have presented an holographic brane-world model able to describe the late-time acceleration of the Universe. The bulk consists of a 5-d Minkowski space-time with a Gauss-Bonnet correction to the usual Hilbert-Einstein action while the brane contains the standard matter content and an holographic energy density together with an induced gravity term. We have assumed that the length that plays the role of the infra-red cutoff of the holographic model is $H^{-1}$, where $H$ is the Hubble rate of the brane. Our analysis was restricted to the normal branch because the other branch can undergo a period of self-acceleration even in the absence of any kind of dark energy on the brane.

We have shown that the normal DGP branch with a Gauss-Bonnet modification can undergo a late-time acceleration consistent with the current observations. Such a consistency is due to the Gauss-Bonnet term. Indeed, the normal DGP branch (in the absence of the Gauss-Bonnet term) with the same kind of holographic energy is decelerating \cite{Wu:2007tp}, likewise the same happens in the standard 4d relativistic model.

We have  shown as well the richness of the modified Friedmann equation of such a model. It not only describes a physically consistent brane but also has mathematically appealing solutions where the brane starts its evolution with a sudden singularity and even ends with the same kind of singularity. Notice in this regard that a sudden singularity is much smoother than a big bang singularity or  a big crunch singularity \cite{Shtanov:2002ek,Nojiri:2005sx,Cattoen:2005dx}.

It is well known that the length that plays the role of the infra-red cutoff on an holographic energy density is not unique, we are currently analysing other physical situations with different lengths which we will report soon \cite{preparation}.

\acknowledgments

The authors are grateful to Pedro F. Gonz\'alez-D\'{\i}az for very useful comments on
a previous version of the manuscript and to Antonio Vale for a careful reading of it.
The authors are as well grateful to a referee for pointing out a mistaken statment on a previous version of the paper. M.B.L. is  supported by the Portuguese Agency Funda\c{c}\~{a}o para a Ci\^{e}ncia e Tecnologia through the fellowship SFRH/BPD/26542/2006. M.B.L. acknowledges the hospitality of the University of Oujda where this work was initiated. A.E. and  T.O. are supported by CNRST, through the fellowship URAC 07/214410.


\begin{thebibliography}{99}
\bibitem{footnote1}
As should be, this is in full agreement with the analysis presented in Sect.~III (cf. Fig.~\ref{pless14}). The condition $16\Omega_{r_c}\Omega_\alpha<(c^2-1)^2$ is equivalent to $b/\bar{c}^2<1/4$, where the parameters $b$ and $\bar c$ are defined in Eqs.~(\ref{defb}) and (\ref{deftildec}).
\bibitem{footnote3} Where now the angle $\theta$  is such that $0\leq\theta<\pi$ where at $\theta=0$, $\bar\rho=\bar\rho_2$.
\bibitem{footnote4}
We can use a similar procedure to the one used in Fig.~\ref{pbcc} to plot $z_1$, the redshift at which the energy density $\bar\rho_1$ is reached [see Eqs.~(\ref{dimensionlessrho}) and (\ref{defbarrho1})] and where the branes with Hubble rates $\bar{H}_2$ and $\bar{H}_3$ are created. We omitted this plot as it is not very enlightening aside from the information we have already explained in the main text.
\end{thebibliography}
\end{document}